\documentclass[]{aa}
\pdfoutput=1
\usepackage{graphicx}
\usepackage{amsmath,amsfonts,amssymb,tabu}
\usepackage[para]{threeparttable}
\usepackage{txfonts}
\usepackage[breaklinks,colorlinks,citecolor=blue,pdfa=true]{hyperref}
\usepackage{color}
\usepackage{fixltx2e}
\usepackage{natbib,twoopt}
\usepackage{url}
\usepackage{multirow}
\usepackage{epsf}
\usepackage{epsfig}
\usepackage{longtable}
\usepackage{float}
\usepackage{subfig}
\usepackage{caption}
\newcommand{\GG}[1]{}

\definecolor{mypink1}{RGB}{219, 48, 122}

\usepackage{ifthen}
\usepackage[T1]{fontenc}
\usepackage{lmodern}
\usepackage{ifxetex,ifluatex}
\usepackage{latexsym}
\usepackage{pdfpages}
\usepackage{dblfloatfix}
\usepackage{morefloats}
\usepackage{caption}
\usepackage{lscape}
\usepackage{mathtools}

\bibpunct{(}{)}{;}{a}{}{,} %% natbib format for A&A and ApJ
\makeatletter
\newcommandtwoopt{\citeads}[3][][]{\href{http://adsabs.harvard.edu/abs/#3}%
{\def\hyper@linkstart##1##2{}%
\let\hyper@linkend\@empty\citealp[#1][#2]{#3}}}
\newcommandtwoopt{\citepads}[3][][]{\href{http://adsabs.harvard.edu/abs/#3}%
{\def\hyper@linkstart##1##2{}%
\let\hyper@linkend\@empty\citep[#1][#2]{#3}}}
\newcommandtwoopt{\citetads}[3][][]{\href{http://adsabs.harvard.edu/abs/#3}%
{\def\hyper@linkstart##1##2{}%
\let\hyper@linkend\@empty\citet[#1][#2]{#3}}}
\newcommandtwoopt{\citeyearads}[3][][]%
{\href{http://adsabs.harvard.edu/abs/#3}
{\def\hyper@linkstart##1##2{}%
\let\hyper@linkend\@empty\citeyear[#1][#2]{#3}}}
\makeatother

\usepackage{graphicx}
\def\mbh   {M$_{\rm BH}$ } 
\def\msun  {M$_{\odot}$ } 
\def\simi   {$\sim$\,}  
\def\z     {$z$ } 
\def\qj    {Q$_{\rm Jet}$ } 
\def\wz    {W Hz$^{-1}$ } 
 
\def\P     {P$_{1400}$ } 
\def\t     {$\times$ } 
\def\pc    {P$_{\rm core}$ }
\def\si    {$\alpha_{150}^{1400}$ }
\def\er    {$\lambda_{\rm Edd}$ }
\def\lb    {L$_{\rm bol}$ }

\begin{document}

   \title{SAGAN - III: New insights into giant radio quasars}

   \titlerunning{SAGAN-III}

\author {Mousumi Mahato\inst{1}\thanks{E-mail: mousumi@iucaa.in}
\and Pratik Dabhade\inst{2,1}
\and D. J. Saikia\inst{1,3}
\and Fran\c{c}oise Combes\inst{2}
\and Joydeep Bagchi,\inst{1,4}\\
L. C. Ho\inst{5,6}
\and Somak Raychaudhury\inst{1,7}
}
  
\authorrunning{Mahato et al.}
\institute{$^{1}$Inter-University Centre for Astronomy and Astrophysics (IUCAA), Pune 411007, India\\ 
$^{2}$Observatoire de Paris, LERMA, Coll\`ege de France, CNRS, PSL University, Sorbonne University, 75014, Paris, France\\
$^{3}$Department of Physics, Tezpur University, Tezpur 784028, India\\
$^{4}$Department of Physics and Electronics, CHRIST (Deemed to be University), Bengaluru-560029, India\\
$^{5}$Kavli Institute for Astronomy and Astrophysics, Peking University, Beijing 100871, People's Republic of China\\
$^{6}$Department of Astronomy, School of Physics, Peking University, Beijing 100871, People's Republic of China\\
$^{7}$School of Physics and Astronomy, University of Birmingham, Birmingham B15~2TT, UK\\ 
}

% \abstract{}{}{}{}{} 
% 5 {} token are mandatory
 \date{\today} 
 
 \abstract
%   % context heading (optional)
%   % {} leave it empty if necessary  
{Giant radio quasars (GRQs) are radio-loud active galactic nuclei (AGNs), propelling megaparsec-scale jets. 
In order to understand GRQs and their properties, we have compiled all known GRQs ("the GRQ catalogue"), and a subset of small (size $<700$ kpc) radio quasars (SRQs) from the literature. In this process, we have found 10 new FR-II GRQs,
in the redshift range of $0.66 < z < 1.72$, which we include in the GRQ catalogue. Using the above samples, we have carried out a systematic comparative study of GRQs and SRQs, using optical and radio data. Our results show that the GRQs and SRQs statistically have similar spectral index and black hole mass distributions. However, SRQs have higher radio core power, core dominance factor, total radio power, jet kinetic power and Eddington ratio compared to GRQs. On the other hand, when compared to giant radio galaxies (GRGs), GRQs have higher black hole mass and Eddington ratio.  The high core dominance factor of SRQs is an indicator of them lying closer to the line of sight than GRQs. We also find a correlation of the accretion disc luminosity with the radio core and jet power of GRQs, which provides evidence for disc-jet coupling. Lastly, we find the distributions of Eddington ratios of GRGs and GRQs to be bi-modal, similar to  that found in small radio galaxies (SRGs) and SRQs, which indicate that size is not strongly dependent on the accretion state. Using all of these, we provide a basic model for the growth of SRQs to GRQs.}


\maketitle

 \section{Introduction} \label{sec:intro}
 Ever since the discovery of quasars \citep{Hazard1963,Schmidt63}, they have been the subject of numerous studies ranging from black hole physics to being probes of large scale structures. Quasars are the most extreme forms of active galactic nuclei (AGNs), and also show radio-loud - radio-quiet dichotomy \citep{Cirasuolo03,Jiang07,Kellermann16,Beaklini20}. The first quasars discovered were of radio-loud nature, and later radio-quiet quasars \citep{sandage65} too were found. The quest to understand the reasons for the radio-loud - radio-quiet dichotomy has been the focus of several studies in past decades, and recent studies \citep{edwinHuub17} suggest that radio-loud quasars (RQs or RLQs) tend to reside in more massive host halos as compared to radio-quiet quasars (RQQs).
 
 Several sample studies \citep[e.g.][]{Hintzen83,Bridle84,Owen84,Bridle94} have been carried out to study and understand the radio morphology of RQs. The extended radio morphology of RQs are mostly of Fanaroff-Riley type-II (FR-II) nature \citep{FR74}, which shows bright emissions in hotspots on either side of the radio core fed by the radio jets. Owing to the bright emission exhibited by the RQs, it is possible to detect high redshift RQs as well as map their radio morphology in detail. 
 
 Since the 1950s, thousands of radio galaxies (RGs) and RQs have been found with a wide range of  sizes and shapes; however, sources with overall sizes greater than 700 kiloparsec (kpc) are found to be relatively rare. Such sources are broadly called giant radio galaxies (GRGs), and if they are powered by a quasar then they are specifically referred to as giant radio quasars or GRQs. Recent studies by \citet{PDLOTSS} and \citet{DabhadeSAGAN20} have shown that only \simi 20\% of the total known GRG population are associated with quasars or GRQs.
 
 The formation and growth of RGs and RQs to megaparsec scales are not completely understood and has been the subject of some recent studies \citep{grscat,PDLOTSS,Kuzmicz2019,DabhadeSAGAN20,sagan2,Bassani21}, which have helped improve our understanding of these relatively rare gigantic objects.
 
 Under the project SAGAN\footnote{\href{https://sites.google.com/site/anantasakyatta/}{Search \& Analysis of Giant radio galaxies with Associated Nuclei}}, we have compiled the list of all known GRGs from the literature along with our newly reported GRGs in paper-I (\citealt{DabhadeSAGAN20}; henceforth SAGAN.I) to study their properties. We also compared their AGN and large scale radio properties with small (size < 700 kpc) RGs (SRGs) in an attempt to identify similarities/dissimilarities between the two. SAGAN.I primarily focused on GRGs with non-quasar AGN and their properties. In this paper (SAGAN.III), we have first compiled a sample of GRQs, which includes 10 new GRQs we report in this paper, 121 GRQs from our catalogue from SAGAN.I and 134 new ones reported by \citet{Kuzmicz2021} (henceforth KJ21) to form a `GRQ-catalogue' consisting of 265 GRQs (Sec.\ \ref{sampecreation}). For this sample, we estimate and consider several physical properties for comparison with smaller sized quasars, as well as with galaxies, to provide insights towards understanding the formation and evolution of these giant sources. 
 
 Recently, KJ21 considered a sample of 272 GRQs and 367 SRQs (small radio quasars; 0.2 $\leq$ size < 0.7 Mpc) and found no evidence of significant differences between the GRQs and SRQs with respect to their optical and infrared properties. The optical properties include black hole mass, Eddington ratio, and distribution of GRQs and SRQs on the Eigenvector 1 plane i.e., an optical plane characterised by the ratio between equivalent width (EW) of FeII and H$\beta$ broad line as a function of FWHM of H$\beta$ broad line. Based on their findings they argue that GRQs and SRQs are evolved AGNs with high black hole mass and low accretion rate. The basic conclusion is similar to their earlier paper from a much smaller sample of sources \citep{Kuzmicz}, where they reported that while GRQs and SRQs differ in size, their black hole mass, accretion rate and prominence of radio cores are similar.

In this paper, we report the finding of 10 new GRQs and present our robust catalogues of GRQs and SRQs. Using these catalogues we have carried out a comparative study of properties of GRQs and SRQs with redshift matched sub-samples. It is important to consider samples matched in redshift to minimize any effects of the evolution of source properties with cosmic epoch. The properties studied here probe parsec to hundreds of kiloparsec scale aspects of the sources using optical and radio data, like the black hole mass, Eddington ratio on a smaller scale to radio power, core dominance factor, jet kinetic power, and integrated spectral index on larger scales. We report significant differences between GRQs and SRQs from the redshift-matched samples. Utilising our results, we connect them to proposed models in literature and interpret our findings in the discussion section, e.g., we explore the prevalence of disc-jet coupling in GRQs and SRQs by showing correlations between AGN bolometric luminosity and jet kinetic power; and between Eddington ratio and Eddington luminosity normalised radio core luminosity. We also compare the black hole mass and Eddington ratio for GRQs and SRQs with giant radio galaxies and smaller sized galaxies to explore differences between galaxies and quasars. Lastly, we also propose a possible model for the formation of GRQs from SRQs based on our findings in this paper and discuss the importance of identifying and studying high redshift GRQs.

 The paper is organised as follows: In Sec.\ \ref{sampecreation}, we describe the methodology of finding the new GRQ sample, creating the GRQ catalogue and the SRQ sample. Sec.\ \ref{Analysis} discusses the multi-wavelength properties of GRQs and SRQs.  In Sec.\ \ref{sec:prop} and the subsections therein, we discuss the methods of analysing the samples in order to carry out a systematic comparative study followed by our results. The Sec.\ \ref{Discussion} outlines the implications of our results and the future prospects, concluded by Sec.\ \ref{Summary} with a brief summary of our results.

The flat $\Lambda$CDM cosmological model has been adopted throughout this paper, based on the Planck results (H$\rm _0$ = 67.8 km s$^{-1}$ Mpc$^{-1}$, $\Omega\rm _m$ = 0.308  and $\Omega\rm _{\Lambda}$ = 0.692; \citealt{2016A&A...594A..13P}). The images are presented in J2000 coordinate system. We have used the  $\rm S_{\nu}\propto \nu^{-\rm \alpha}$ convention for the rest of the paper, where S$_{\nu}$ is the flux density at frequency $\nu$ and $\rm \alpha$ is the spectral index.

%%%%%%%%%%%%%%%%%%%%%%%%%%%%%%%%%%%%%%%%%%%%%%%%%%%%%%%%%%%%%%%%
\begin{figure*}
\centering
\includegraphics[scale=0.21]{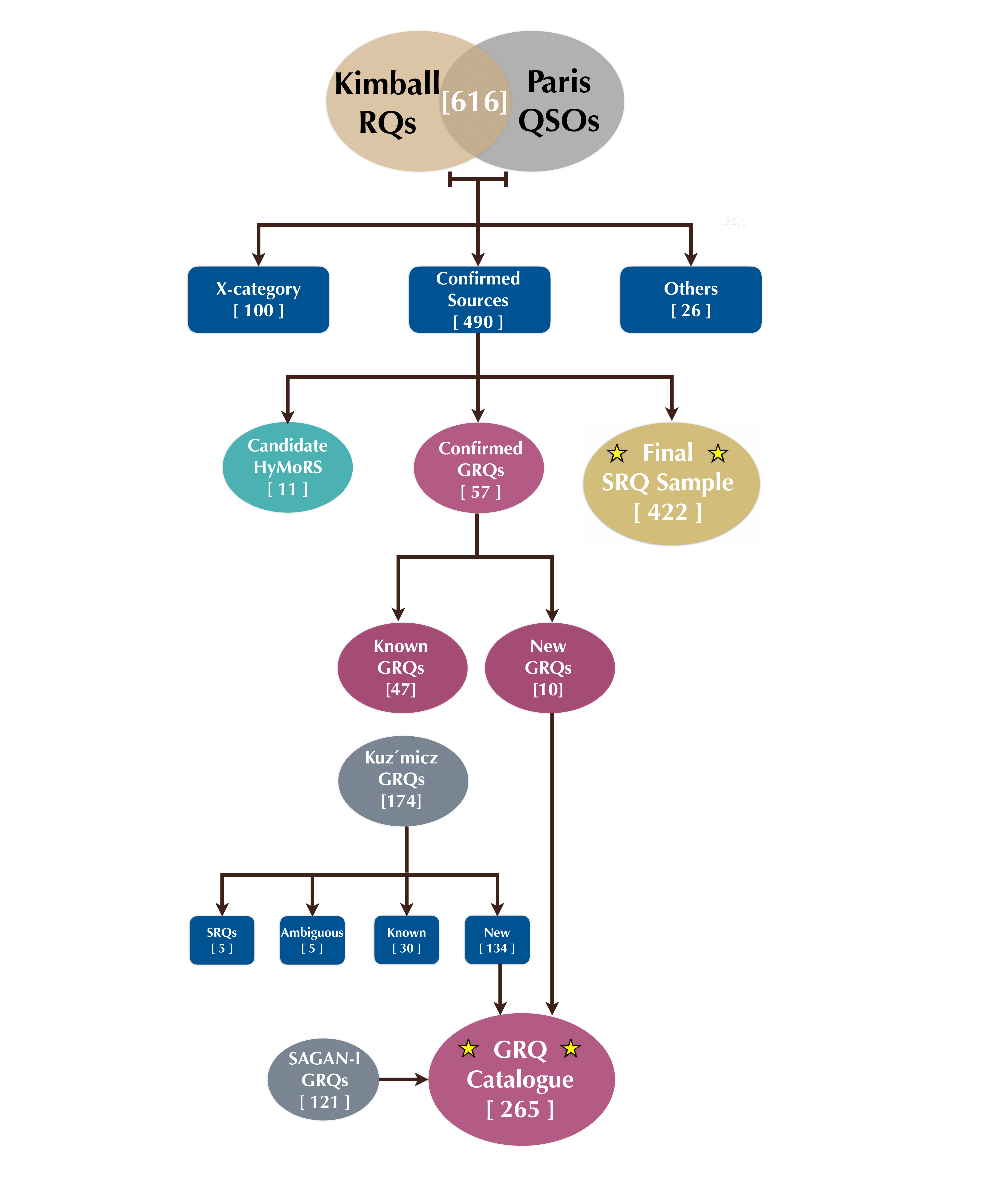}
\caption{Flow chart for creating the GRQ catalogue and the SRQ sample used for our analysis in this paper. Here, Kimball refers to the RQ catalogue of \citet{Kimball}, Paris to the Quasar catalogue of \citet{parisqso}, Ku\'zmicz to the GRQ catalogue by \citet{Kuzmicz2021}, and SAGAN-I to the GRG catalogue of \citet{DabhadeSAGAN20}. The 'Others' category consists of 19 sources with ambiguous morphology, 5 sources showing ghost emission due to artefacts, and 2 with false detection. The details of the steps illustrated here can be found in Sec.\ \ref{sampecreation}.}
\label{fig:flow}
\end{figure*} 
%%%%%%%%%%%%%%%%%%%%%%%%%%%%%%%%%%%%%%%%%%%%%%%%%%%%%%%%%
 
\begin{figure*}
\centering
\includegraphics[scale=0.35]{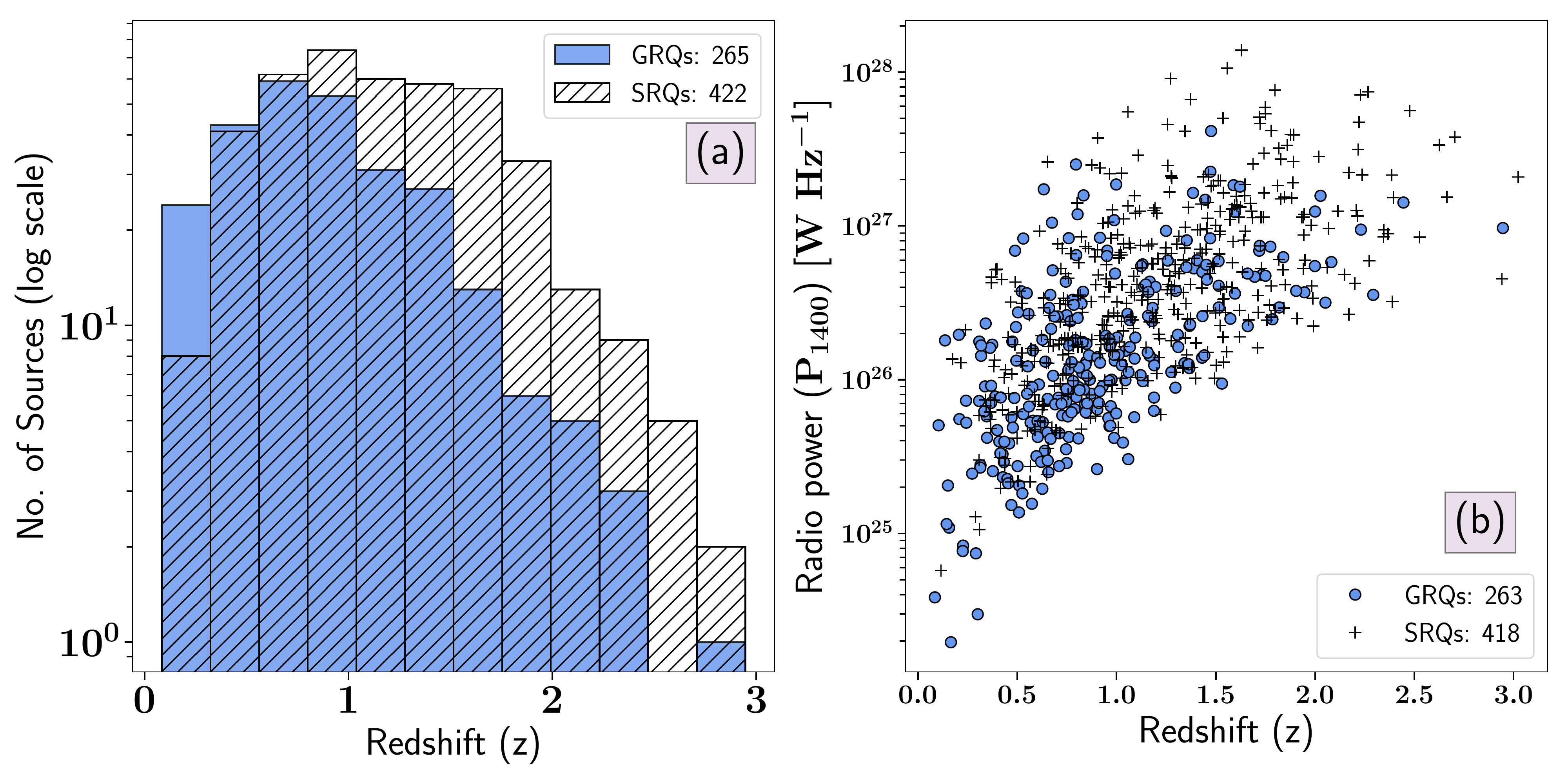}
\caption{Left (a): The redshift ($z$) distributions of all GRQs and SRQs. Right (b): All the GRQs and SRQs are plotted on the $P\!-\!z$ plane where the blue circles represent the GRQs and the '+' sign the SRQs. The 1400~MHz NVSS flux density measurements were not available for two GRQs due to their position beyond the coverage area of the NVSS. For four SRQs, the flux densities at 1400 MHz are contaminated by the presence of other sources due to the low resolution of the NVSS. Therefore, (b) has 418 out of 422 SRQs, and 263 out of 265 GRQs.}
\label{fig:z_Pz}
\end{figure*} 

%%%%%%%%%%%%%%%%%%%%%%%%%%%%%%%%%%%%%%%%%%%%%%%%%%%%%%%%%%

\section{Sample}\label{sampecreation}
In this section we describe the details of GRQ and SRQ samples. Two subsections are dedicated to GRQs and one for SRQs.
\subsection{The new GRQ sample}
\label{GRQ_Samp}
\citet{Kimball} (hereinafter K11) have created a large sample of radio quasars with optical data from the Sloan Digital Sky Survey (SDSS; \citealt{sdssyork}) and the radio data from the Faint Images of the Radio Sky at Twenty-centimeters (FIRST; \citealt{beckerfirst95}). They have used an automated algorithm to classify the radio morphology of the sources. The entire K11 sample of 4714 objects has radio flux density greater than 2 mJy at 1400 MHz. Out of them, 619 sources are reported to have proper lobe-core-lobe (triple) radio morphology. This compilation of 619 objects serves as a useful database to create an SRQ sample and to find new GRQs. Hence, we have carried out an independent visual inspection of 619 RQs using radio surveys like the FIRST, NRAO VLA Sky Survey (NVSS; \citealt{nvss}), TIFR GMRT Sky Survey alternative data release-1 (TGSS ADR1; \citealt{intema-tgss-17}), and Very Large Array Sky Survey (VLASS; \citealt{vlass}) to verify their morphology and to check for possible extended emission missed by the FIRST. As a result, we have found 10 new GRQs reported in this paper along with their radio properties (Tab\ \ref{tab:maintab}). The radio images of the GRQs are presented in Fig.\ \ref{fig:im1} and Fig.\ \ref{fig:im2}.

\subsection{The GRQ catalogue}
The GRQ catalogue is the compendium of all the GRQs reported in the literature and our new GRQ sample. It consists of 265 sources out of which 121 are from the GRG catalogue in the SAGAN.I paper \citep{DabhadeSAGAN20}, 134 are from KJ21 and 10 are the new GRQs reported in this paper. We have remeasured all angular sizes from hotspot to hotspot uniformly to ensure that our classification as a giant source is robust, as described in more detail in Sec.\ \ref{size}. Although KJ21 reported 174 new GRQs, upon careful examination we found that 30 of these had already been reported earlier in the literature \citep{Amirkhanyan2016,hardcastle-hatlas,koziel11}. We have not included 10 sources, 3 (J0034+3330, J1014+6047, and J1432+5200) of which require further radio observations to clarify their radio structures, including a possible association of nearby components, in order to get reliable values of the source parameters. A further two (J0846+1413 and J1408+1010) were excluded as these did not occur in the catalogue of confirmed quasars from the SDSS \citep{parisqso}. These 5 have been mentioned as `ambiguous' in Fig.\ \ref{fig:flow} which depicts how the GRQ catalogue has been made. The rest 5 sources (J0053-0210, J0235+0329, J0811+1652, J0849+4216 and J1145+4423) are found to be slightly smaller than our limit of 700 kpc and hence, not included in our sample of GRQs. With the recent emphasis and finding of new GRQs, the GRQ catalogue now constitutes \simi30\% of the total GRG (quasar + non-quasar AGN) population known to date.

\subsection{The SRQ sample}
In order to compare the properties of GRQs with SRQs, one needs to create a robust SRQ catalogue (size $<$ 700\,kpc). Hence, we have compiled a sample of SRQs from the catalogue of K11, as described in Sec.\ \ref{GRQ_Samp}. The K11 catalogue was compiled using optical data from the SDSS DR5 and the ancillary FIRST radio data. Hence, it is essential to update the redshifts from the latest products of SDSS (DR16), which have been revised over the years. We cross-matched the `triple' category of 619 sources from the K11 catalogue with the robust quasar catalogue of \citet{parisqso}. As a result, we have found a total of 616 RQs to have reliable redshifts from \citet{parisqso}. Of the 616 RQs, 100 sources belong to the `unclassified' or `X' category as given by K11. Based on our analysis, we have identified 19 sources to have ambiguous radio morphology, 5 sources have ghost emission (i.e., artefacts), 2 sources are wrongly classified and 11 are found to be candidate Hybrid Morphology Radio Sources (HyMoRS; \citealt{saikia-hymors}; \citealt{GKhymorph}). These sources (100+19+5+2+11) are not being considered further in our study. However, we have found 57 (47 are known, 10 are new) GRQs from the remaining 479 sources, and hence, the final SRQ sample of 422 sources is formed. The above steps are illustrated in form of a flow chart as seen in the Fig.\ \ref{fig:flow}.

\section{Analysis}
\label{Analysis}
In this section we describe the methods used for computing various properties of GRQ and SRQ samples.

 \subsection{Size}
 \label{size}
\begin{figure*}
\centering
\includegraphics[scale=0.50]{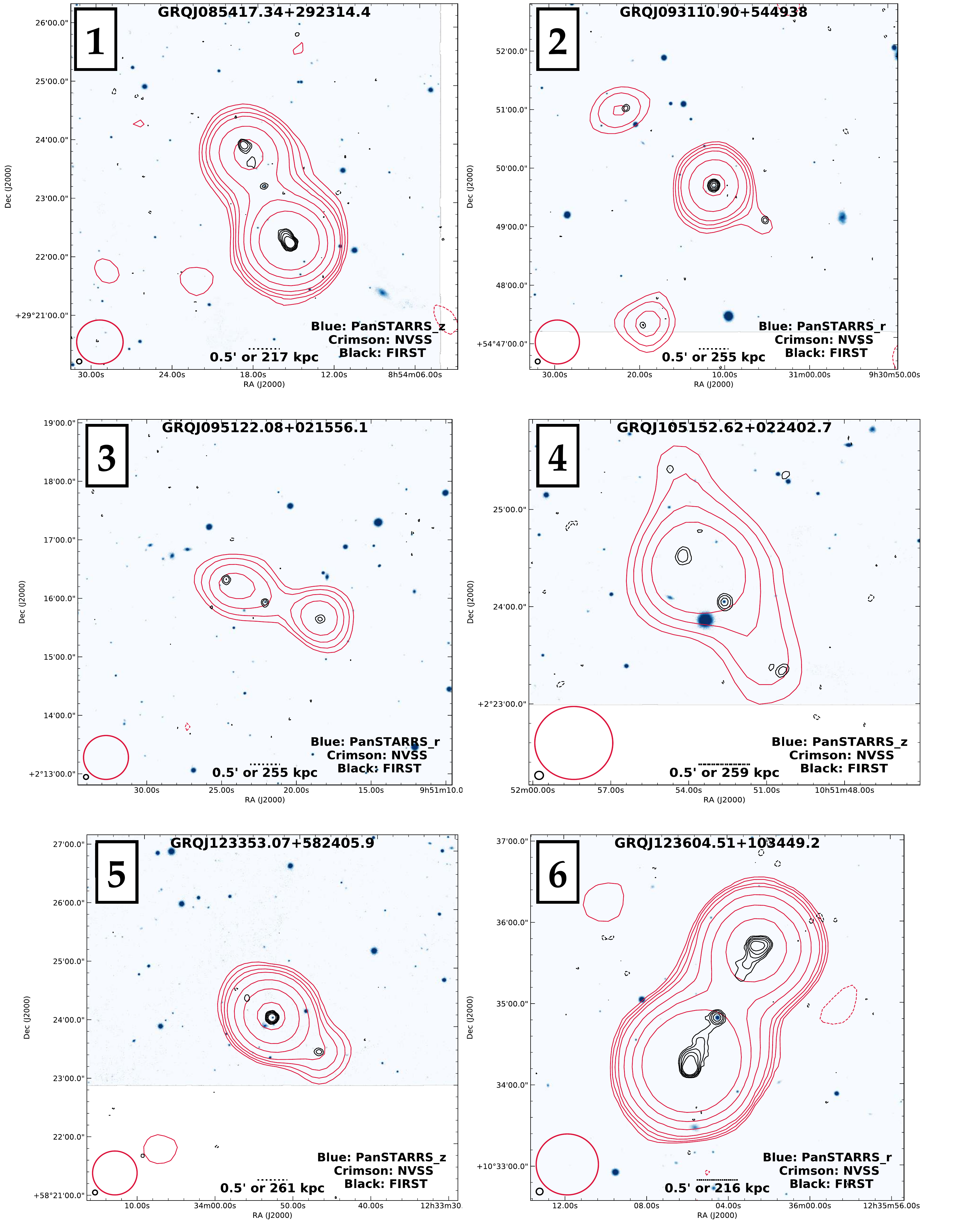}
\caption{Here we present the radio maps of the 10 new GRQs (this page has numbers 1 to 6) as described in Sec.\ \ref{GRQ_Samp}. Here the crimson colour contour represents the radio emission detected in the NVSS, with black contours from the FIRST survey. The background (blue) colour images are from the PanSTARRS $r$-band. In the lower-left corner, the beams of the NVSS and FIRST are shown. The bottom middle of each image shows angular scale for reference. The numbers on the upper left corner represent the serial numbers from Table.\ \ref{tab:maintab}. Contour levels are at 3$\sigma~\times$\,[-1,1,1.4,2.0,2.8,5.6,11.2,22.4]
for NVSS and 3$\sigma~\times$\,[-1,1.4,2.0,2.8,5.6,11.2,22.4] for FIRST, where dashed contours represent negative values. The typical $\sigma$ values for the FIRST and NVSS are 0.15 and 0.45 mJy~beam$^{-1}$, respectively.}
\label{fig:im1}
\end{figure*}

\begin{figure*}
\centering
\includegraphics[scale=0.50]{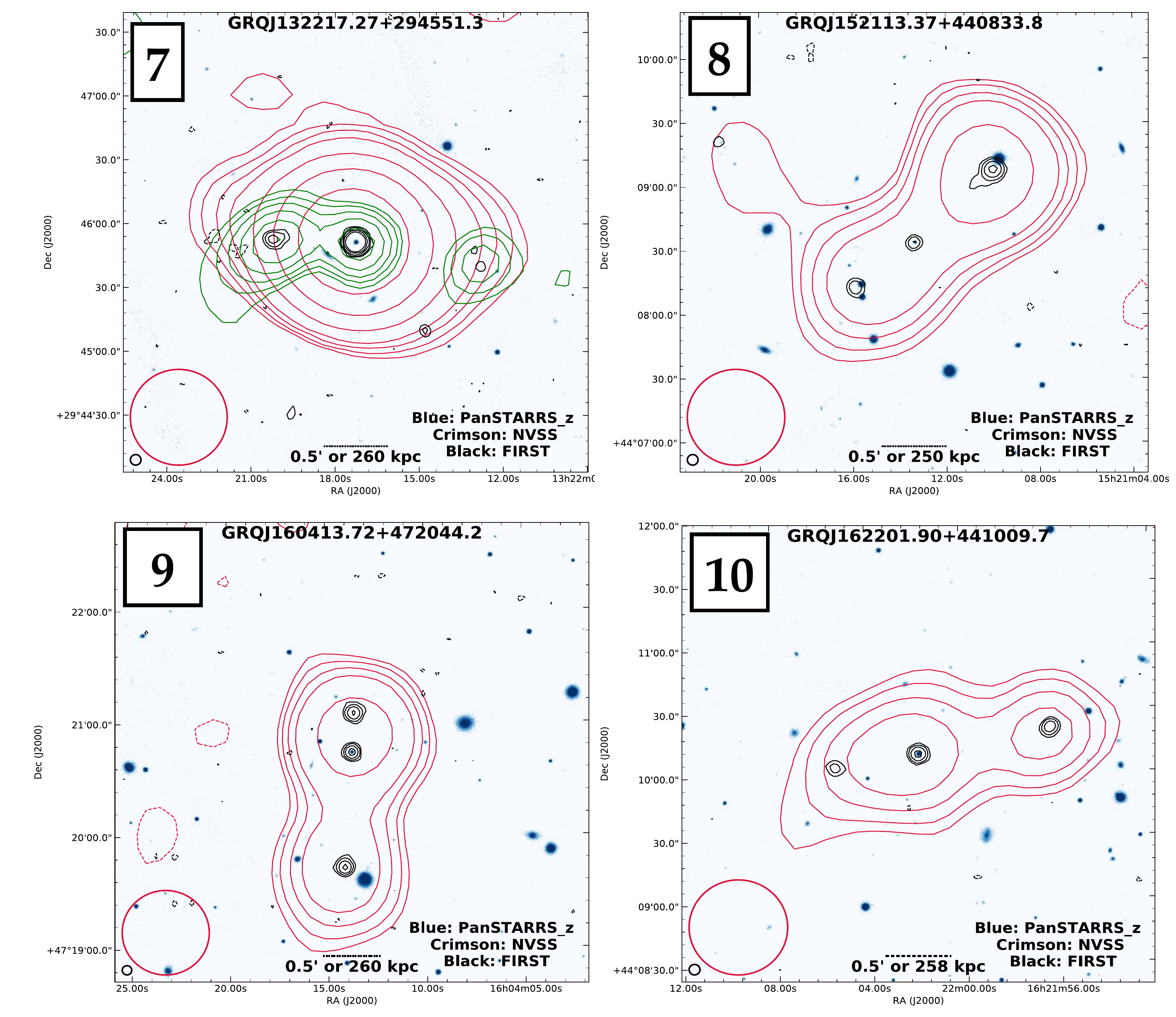}
\caption{The same as in Fig.\ \ref{fig:im1}, but for GRQ number 7--10, as described in Sec.\ \ref{GRQ_Samp}. In the case of GRQ 7 or GRQJ132217.27+294551.3, the radio emission from the TGSS is shown in green contours for clarity of the source morphology, where the $\sigma$ is 2.7~mJy~beam$^{-1}$ and six contours levels are drawn above 3$\sigma$.}
\label{fig:im2}
\end{figure*}

 All the new 10 GRQs reported in this paper show FR-II radio morphology as seen in the radio maps (Fig.\ \ref{fig:im1} and Fig.\ \ref{fig:im2}). The projected linear size has been measured as the distance between two hotspots (peak flux densities within 3 sigma contours) of lobes. The sizes are measured using the formula given in section 3 of SAGAN.I paper. For the radio quasars in the SRQ sample, the FIRST radio maps have been used for the measurement and confirmation of sizes. In case of the sources in the GRQ catalogue, the angular sizes of 176 GRQs have been estimated from the FIRST radio maps. For 17 GRQs, the angular sizes are taken from their respective reporting papers \citep{3crr,Bruyn1989,BhatnagarGRQ,Ishwara1999,Schoenmakerspart2,Schoenmakerspart1,lara2001,Machalski2001,Saripalli2005,Machalski2007b,hardcastle-hatlas} as they had better maps than FIRST. The sizes of the rest 72 sources are measured using the LOFAR Two-meter Sky Survey (LoTSS; \citealt{shimwell-lotss}; 18), TGSS ADR1 (25) and NVSS (29) radio images due to unavailability of the FIRST radio maps. The same method, as described above, has been employed for all size measurements. All the radio sources considered in this paper belong to the FR-II class. 
 
%%%%%%%%%%%%%%%%%%%%%%%%%%%%%%%%%%%%%%%%%%%%%%%%%%%

%%%%%%%%%%%%%%%%%%%%%%%%%%%%%%%%%%%%%%%%%%%%%%%%%%%

 \subsection{Radio core power}
 The radio core power (P$_{\rm core}$) of the SRQ sample is estimated at the rest-frame frequency of 1400 MHz from the observed core flux densities at 1400 MHz (FIRST) taken from K11 using the spectral index of 0 (\citealt{Bridle94}; K11; \citealt{2020Maithil}). For sources in the GRQ catalogue, the core flux densities
 are obtained from the FIRST source catalogue \citep{Helfand15}.

 \subsection {Total radio power}
 The integrated flux densities (entire source) at 150 MHz and 1400 MHz of the SRQs and the GRQs are measured from the TGSS and the NVSS radio maps, respectively, using CASA\footnote{Common Astronomy Software Applications (CASA; \citealt{casa}) task `CASA-VIEWER'.} by selecting the regions of radio emission of the sources manually. For estimating the errors in the flux density measurements, we employed the method of \citet{klein03} considering 20\% and 3\% flux density calibration errors for the TGSS and NVSS, respectively. The total radio power at the respective rest-frame frequencies (P$_{150}$ and P$_{1400}$) are tabulated using the formula given in section 3 of SAGAN.I paper.

%%%%%%%%%%%%%%%%%%%%%%%%%%%%%%%%%%%%%%%%%%%%%%%%%%%%

\subsection{Core dominance factor}
 The core dominance factor (CDF) is the ratio of core flux density ($\rm S_{core}$) to the extended flux density ($\rm S_{ext}$). The extended flux density is the flux density from the extended regions of the RG/RQ except from the core (e.g., lobes). It is estimated by subtracting the core flux density (measured from the FIRST) from the total integrated flux density (measured from the NVSS). The CDF has been estimated at the rest-frame frequency ($\rm \nu_{rest}$) of 5000~MHz (or 5~GHz) for the sources. We have derived the flux densities (both $\rm S_{core}$ and $\rm S_{ext}$) at the 5 GHz rest-frame frequency from the respective flux densities observed at 1400 MHz ($\rm \nu_{obs}$) using the following formula:
 \begin{equation}
   \rm  S_{\nu_{rest}} = S_{\nu_{obs}}(1+z)^{(\alpha -1)}\left(\frac{\nu_{rest}}{\nu_{obs}}\right)^{(-\alpha)}
 \end{equation}
 where $\rm S_{\nu_{rest}}$ is the rest-frame flux density at $\rm \nu_{rest}$, and $\rm S_{\nu_{obs}}$ is the observed flux density at $\rm \nu_{obs}$ of the core and extended regions. The redshift is denoted by $z$ for the source. Here, the spectral index $\alpha$ is assumed to be 0 and 0.75 \citep{thorpe99,2020Maithil} for the radio core and the extended emission, respectively. The CDF has also been referred to as `core fraction' (f$_{c}$), `core prominence', and `radio core dominance' in the literature. It is also often denoted as `R'. 
 
 %%%%%%%%%%%%%%%%%%%%%%%%%%%%%%%%%%%%%%%%%%%%%%%%%%%

 \subsection{Jet kinetic power}
 The jet kinetic power (Q$_{\rm Jet}$) is estimated from the low-frequency radio observations to avoid the effects of Doppler boosting, which is prominent in higher-frequency observations. For our analysis, it is measured using total radio power at 150~MHz from the TGSS maps, as described in section 3 of SAGAN.I. For the GRQs in the GRQ catalogue taken from \citet{PDLOTSS}, the 144 MHz flux densities from the LoTSS are used to estimate the jet kinetic power.

\subsection{Spectral Index}
The two-point spectral index ($\alpha_{150}^{1400}$) is computed using the flux densities measured at 150 MHz (TGSS) and 1400 MHz (NVSS). For the sources which are either partially or not detected in TGSS, we have considered spectral index to be 0.70 for both GRQs and SRQs. This is the median spectral index value of both the GRQ catalogue and the SRQ sample, which is used further for the estimation of the total radio powers.

\subsection{Black hole mass}
\label{MBH}
 We have cross-matched our GRQ catalogue and the SRQ sample with the SDSS-DR14 quasar catalogue made by \citet{Rakshit2020} to obtain the black hole masses (M$_{\rm BH}$), estimated using spectroscopic emission lines. They have used the emission lines of $\rm H\beta$, MgII and CIV for the redshift range of \z $<$ 0.8, 0.8 $\leq$ \z $<$ 1.9, \z $\geq$ 1.9, respectively. The black hole masses for 33 GRQs and 1 SRQ are not available from the catalogue of \citet{Rakshit2020}. The M$_{\rm BH}$ of 5 \t 10$^{7}$ \msun corresponds to stellar velocity dispersion ($\sigma$) of 70~km s$^{-1}$, which is the instrumental resolution limit of the SDSS spectra. Hence, only sources with masses above 5 \t 10$^{7}$ \msun were considered. We have also taken into account other quality filters recommended by \citet{Rakshit2020} while considering the M$_{\rm BH}$. Hence, 41 GRQs and 49 SRQs were not taken into account for analysing M$_{\rm BH}$ and related properties.

\subsection{Eddington ratio}
The Eddington ratio ($\lambda_{\rm Edd}$) of a source is related to the accretion rate of its supermassive black hole (SMBH). It is defined as the ratio of bolometric luminosity (L$_{\rm bol}$) to the Eddington luminosity (L$_{\rm Edd}$) corresponding to its mass. Similar to M$_{\rm BH}$ as described in Sec.\ \ref{MBH}, L$_{\rm bol}$ is also obtained from the sample of \citet{Rakshit2020}, who have used monochromatic line luminosity at 5100 \AA, 3000 \AA, and 1350 $\AA$ to estimate L$_{\rm bol}$ for the $z$ range of $z$ < 0.8, 0.8 $\leq$ $z$ < 1.9 and $z$ $\geq$ 1.9 respectively. We have derived  L$_{\rm Edd}$ from M$_{\rm BH}$ using the formula: L$_{\rm Edd}$ = 1.3 $\times$ 10$^{38}$ $\times$ ($\frac{\rm M_{\rm BH}}{\rm M_{\odot}}$) erg s$^{-1}$.
 
\section{Results}
\label{sec:prop}
The redshift ($z$) for the SRQ sample (422) ranges from 0.11 to 3.02, and for the GRQs (265), it ranges from 0.08 to 2.94. The redshift distributions of the final GRQ and SRQ samples are shown in Fig.\ \ref{fig:z_Pz} (a). The sample of GRQs has a mean redshift of about 0.91 whereas the distribution of SRQs has a mean value of 1.20. Fig.\ \ref{fig:z_Pz} (b) shows the distribution of GRQs and SRQs on the P-\z (radio power - redshift) plane where the radio power spans over 3 orders of magnitude. The non-availability of sources in the lower right quadrant is due to  Malmquist bias. It is observed that with the increase in redshift the number of sources decreases. Most of the sources are populated around \z \simi0.5 to 1.5. The most powerful GRQs with radio power \simi 10$^{27}$ \wz are clustered between redshift 1.0 to 2.0.

 Our aim is to understand how different or similar are the SRQ and GRQ populations in terms of their optical and radio properties, and under what conditions the SRQs are growing to GRQs. To achieve this, we have studied the samples by comparing properties in redshift-matched sub-samples. 
In this method, we have divided the samples into two redshift bins depending on the median values of the entire \z distributions. The median value of the GRQ sample is 0.82 whereas the median \z value for the SRQ sample is 1.16. We have considered the average of these two median values (\z = 1.00) to divide the GRQ and SRQ sample into two bins. Both the bins i.e., lower (\z $\leq$ 1.00) and upper (1.00 < \z $\leq$ 2.45) are matched in redshift with K-S test p-value and WMW test p-value for lower bin being 0.23 and 0.05, and for upper bin being 0.11 and 0.05, respectively. Fig.\ \ref{fig:zmed1} (a) and Fig.\ \ref{fig:zmed2} (a) show the respective distributions. 

Based on the above method, we have compared properties like P$_{\rm core}$, CDF, P$_{1400}$, Q$_{\rm Jet}$, $\alpha_{150}^{1400}$, \mbh and \er of GRQs and SRQs as discussed in Sec.\ \ref{Analysis}. The results are summarised in Tab.\ \ref{zrpmed}, where we have provided the statistics related to all sub-samples with respect to their properties, like the mean, median, and the p values of the statistical tests. We reject the null hypothesis that the two samples were drawn from the same distribution if the p-value is less than the significance level.

\subsection{Distributions of radio core power}
\label{corepower}
The \pc distributions of GRQs and SRQs are shown in Fig.\ \ref{fig:zmed1} (b) for the lower bin and in Fig.\ \ref{fig:zmed2} (b) for the upper bin of redshift. Both distributions show that the SRQs have higher \pc than GRQs. This is supported by p-values of statistical tests given in Tab.\ \ref{zrpmed}. The \pc of SRQs is found to be more than twice that of GRQs.

%%%%%%%%%%%%%%%%%%%%%%%%%%%%%%%%%%%%%%%%%%%%%%%%%%

 \onecolumn
\setlength{\tabcolsep}{4pt}
\setlength\extrarowheight{3pt}
% \begin{small}
\begin{landscape}
\begin{longtable}{c c c c c c c c c c c c c c }
\captionsetup{width=9.4in}
\caption{Basic radio properties of the 10 new GRQs. Columns (2) $\&$ (3) represent the right ascension (RA) in HMS (hr min s) and declination (Dec) 
in DMS ($^\circ$~\arcmin~\arcsec) of the host of the GRQs. Column (4) lists the spectroscopic redshifts of the hosts. Columns (5) and (6) show the angular size and the projected linear sizes of the sources in arcmins and megaparsecs, respectively. Columns (7) and (9) list the integrated flux densities ($\rm S_{\rm \nu}$), and columns (8) and (10) the total power $\rm P_{\rm \nu}$ of the sources 
at 1400 MHz and 150 MHz, respectively. Column (11) states the $\rm \alpha_{\rm 150}^{\rm 1400}$ which is the two-point spectral index between 1400 MHz (NVSS) and 150 MHz (TGSS). Column (12) contains the jet kinetic power of the sources. Column (13) lists the morphological type of the sources (II representing FR-II type).} \\

\hline 
Sr.No & RA & Dec & z & Size  & Size & $\rm S_{1400}$ & $\rm P_{1400}$ & $\rm S_{150}$ & $\rm P_{150}$ & $\rm \alpha^{\rm 1400}_{\rm 150}$ & Q$_{\rm Jet}$ & Morphology\\
         & (hr min s) & ($^\circ$~\arcmin~\arcsec) &   & (\arcmin) & (Mpc) & (mJy) & ($ 10^{25}$W Hz$^{-1}$) & (mJy)& ($ 10^{25}$W Hz$^{-1}$) &  & ($ 10^{43}$ erg s$^{-1}$)\\
(1) &(2)&(3)&(4)&(5)&(6)&(7)&(8)&(9)&(10)&(11)&(12)&(13)    \\
\hline

1 & 08 54 17.34 & 29 23 14.4 & 0.67075$\pm$0.00007 & 1.81 & 0.78 & 121$\pm$4 & 21.4$\pm$1.2 & - & - & - & - & II \\ 

2 & 09 31 10.90 & 54 49 38.0 & 1.18915$\pm$0.00026 & 3.03 & 1.55 & 20$\pm$1 & 13.4$\pm$1.2 & - & - & - & - & II \\ 

3 & 09 51 22.08 & 02 15 56.1 & 1.18890$\pm$0.00037 & 1.70 & 0.87 & 11$\pm$1 & 7.6$\pm$0.9 & - & - & - & - & II \\ 

4 & 10 51 52.62 & 02 24 02.7 & 1.36687$\pm$0.00065 & 1.51 & 0.78 & 12$\pm$1 & 11.9$\pm$1.5 & 66$\pm$15 & 63.3$\pm$15.3 & 0.75$\pm$0.25 & 21.1$\pm$5.1 & II \\ 

5 & 12 33 53.07 & 58 24 05.9 & 1.51526$\pm$0.00031 & 1.51 & 0.79 & 46$\pm$2 & 29.5$\pm$2.9 & 53$\pm$12 & 34.3$\pm$8.2 & 0.07$\pm$0.23 & 11.4$\pm$2.7 & II \\ 

6 & 12 36 04.51 & 10 34 49.2 & 0.66668$\pm$0.00005 & 1.70 & 0.74 & 221$\pm$7 & 35.6$\pm$1.9 & 749$\pm$151 & 121.3$\pm$24.9 & 0.55$\pm$0.21 & 40.2$\pm$8.3 & II \\ 

7 & 13 22 17.27 & 29 45 51.3 & 1.71883$\pm$0.00066 & 1.61 & 0.84 & 75$\pm$3 & 69.0$\pm$7.0 & 113$\pm$24 & 105.2$\pm$24.4 & 0.19$\pm$0.22 & 34.9$\pm$8.1 & II \\ 

8 & 15 21 13.37 & 44 08 33.8 & 1.06050$\pm$0.00042 & 1.40 & 0.70 & 22$\pm$1 & 11.2$\pm$0.9 & - & - & - & - & II \\ 

9 & 16 04 13.72 & 47 20 44.2 & 1.44235$\pm$0.00052 & 1.37 & 0.71 & 20$\pm$1 & 14.4$\pm$1.7 & 39$\pm$10 & 28.8$\pm$7.9 & 0.31$\pm$0.27 & 9.6$\pm$2.7 & II \\ 

10 & 16 22 01.90 & 44 10 09.7 & 1.29678$\pm$0.00031 & 1.66 & 0.86 & 11$\pm$1 & 8.9$\pm$1.1 & - & - & - & - & II \\ 

\hline
\label{tab:maintab}
\end{longtable}
\end{landscape}
% \end{small}
\twocolumn

%%%%%%%%%%%%%%%%%%%%%%%%%%%%%%%%%%%%%%%%%%%%%%%%%%%

\subsection{Distributions of core dominance factor}
The radio cores of powerful radio-loud AGNs (e.g. quasars) are Doppler boosted due to relativistic beaming \citep{blandford79,Scheuer79,Orr82,kapahisaikia82} if the source axis is close to our line of sight. Hence, the CDF can act as an independent statistical estimator of orientation of the sources \citep{wills95,marin16}.
The CDF distributions of SRQs and GRQs for both redshift matched samples are shown in Fig.\ \ref{fig:zmed1} (c) and Fig.\ \ref{fig:zmed2} (c) for lower and upper bins of redshift, respectively. The range of CDF values of SRQs and GRQs are consistent with the findings of previous studies \citep{Browne87,marin16,2020Maithil}.
We find that the CDF of SRQs is more than GRQs by a factor of nearly two for both redshift matched samples.
This implies that radio cores of SRQs appear more powerful than GRQs which is also seen in Sec.\ \ref{corepower}. It may be relevant to note here that for galaxies alone, which are believed to be inclined at large angles to the line of sight in the unification scheme, \citet{Ishwara1999} did not find any significant difference in the degree of core prominence between giant radio galaxies and smaller ones when they are matched in total radio luminosity.

\subsection{Distributions of total radio power}
The distributions of \P for SRQs and GRQs for redshift matched samples are shown in Fig.\ \ref{fig:zmed1} (d) and Fig.\ \ref{fig:zmed2} (d) for lower and upper bins, respectively. Similar to \pc and CDF, the trend for \P is also the same with SRQs being more powerful than GRQs.

\subsection{Distributions of jet kinetic power}
The \qj shows no exception to the earlier three properties as reflected in Fig.\ \ref{fig:zmed1} (e) and Fig.\ \ref{fig:zmed2} (e) for lower and upper bins of redshift matched samples and statistical test results tabulated in Table\ \ref{zrpmed}. It indicates that SRQs have more powerful jets as compared to GRQs.

\subsection{Distributions of the spectral index}
The \si distributions for GRQs and SRQs are found to be similar for lower as well as upper bins of redshift matched samples. The distributions of \si are shown in Fig.\ \ref{fig:zmed1} (f) for the lower bin and in Fig.\ \ref{fig:zmed2} (f) for the upper bin, respectively. Analogous to SRGs and GRGs \citep{DabhadeSAGAN20}, the SRQs and GRQs have similar spectral index values.

\subsection{Distributions of black hole mass}
\label{mbh}
The distributions of \mbh for redshift matched samples are shown in Fig.\ \ref{fig:zmed1} (g) and in Fig.\ \ref{fig:zmed2} (g) for lower and upper bins, respectively. The \mbh of GRQs is found to be similar to SRQs. This result is in line with the case of GRGs and SRGs \citep{DabhadeSAGAN20} where the black hole masses of GRGs and SRGs have similar distributions peaking at similar values. Moreover, if we compare \mbh of GRGs and SRGs (Table.3, SAGAN.I and Fig.\ \ref{fig:mbher}) with GRQs and SRQs, we notice that the black holes of the latter population are more massive than the former one.

\subsection{Distributions of Eddington ratio}
\begin{figure*}
\centering
\includegraphics[scale=0.24]{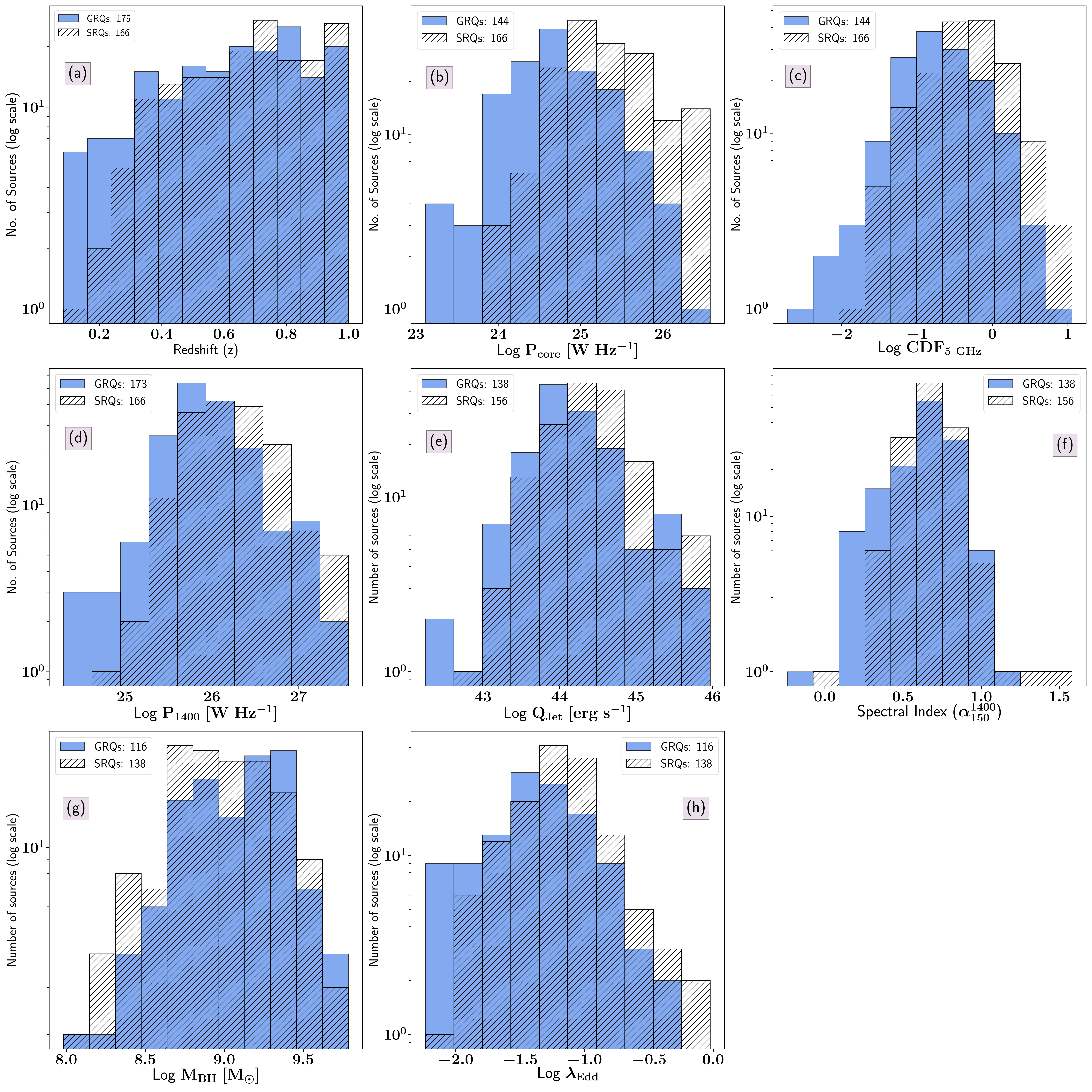}
\caption{This shows the properties of the redshift-matched samples ($z$ $\leq$ 1.00), where GRQs and SRQs are represented in unhatched and hatched bins, respectively. The mean and median values of the distributions are given in Tab.\ \ref{zrpmed}. The panels show (a): redshift distributions for $z$ $\leq$ 1.00;  (b): distributions of radio core power (P$_{\rm core}$) at 1400 MHz obtained from the FIRST; (c): distributions of core dominance factor (CDF); (d): distributions of total radio power (P$_{\rm 1400}$) at 1400 MHz obtained from the NVSS; (e): distributions of jet kinetic power (Q$_{\rm Jet}$); (f): distributions of spectral index ($\alpha_{\rm 150}^{1400}$) estimated using flux densities at 150 MHz and 1400 MHz; (g): distributions of black hole mass (M$_{\rm BH}$); and (h): distributions of Eddington ratio ($\lambda_{\rm Edd}$) of GRQs and SRQs.}
\label{fig:zmed1}
\end{figure*}

\begin{figure*}
\centering
\includegraphics[scale=0.24]{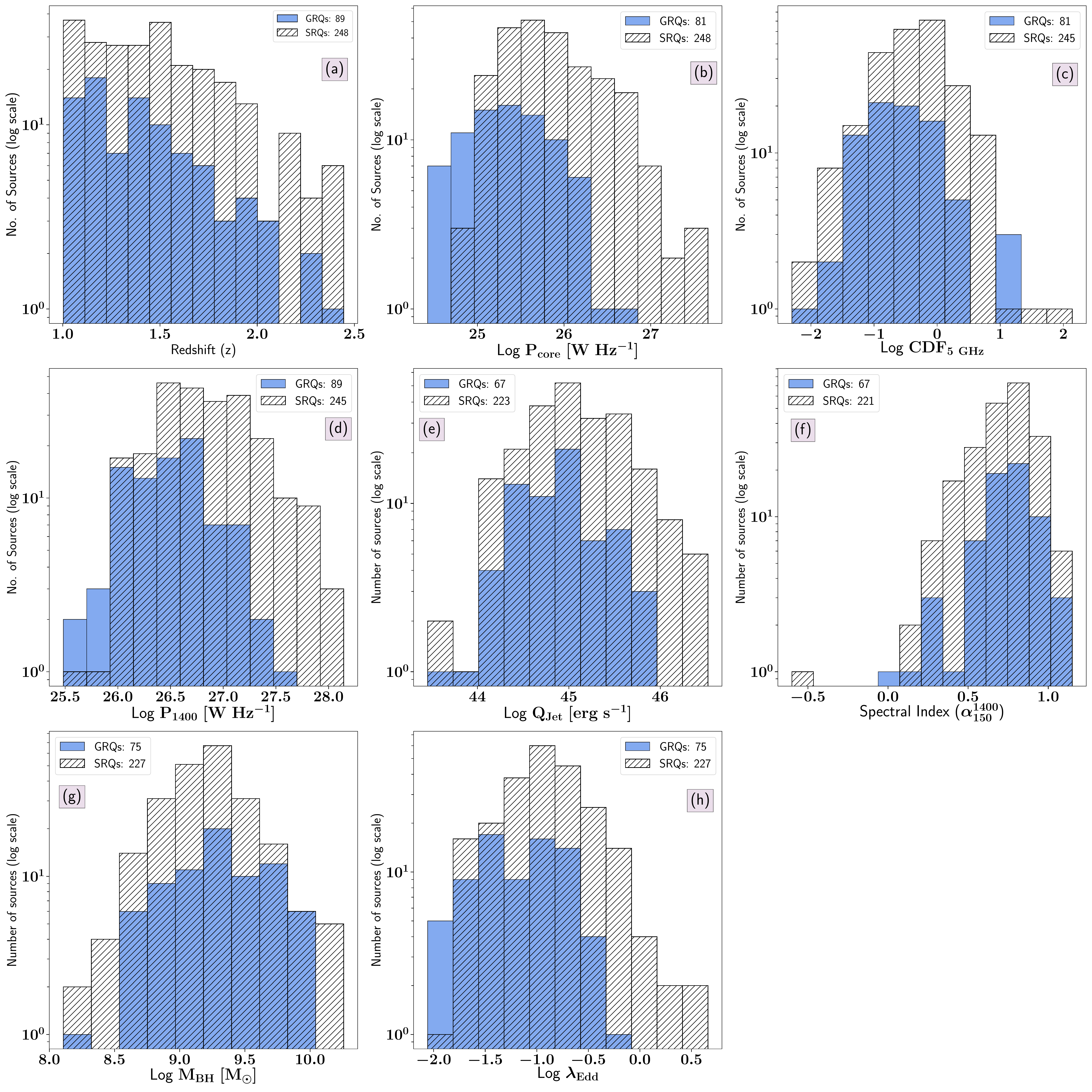}
\caption{Same as Fig.\ \ref{fig:zmed1} for 1.00 < $z$ $\leq$ 2.45.}
\label{fig:zmed2}
\end{figure*}

The results of comparing the distributions of Eddington ratios of GRQs and SRQs are found to be on similar lines with that of the GRGs and SRGs \citep{DabhadeSAGAN20}. This accounts for the fact that SRQs have higher \er than GRQs like SRGs when compared to GRGs. The \er distributions for SRQs and GRQs can be seen in Fig.\ \ref{fig:zmed1} (h) and Fig.\ \ref{fig:zmed2} (h) for lower and upper bins of redshift matched samples, respectively. When we compare GRGs and SRGs with GRQs and SRQs, it is interesting to find that \er of the galaxy population is less than the quasar population (Table.3, SAGAN.I and Fig.\ \ref{fig:mbher}).

%%%%%%%%%%%%%%%%%%%%%%%%%%%%%%%%%%%%%%%%%%%%%%%%%%%%

\begin{table*}
\setlength \extrarowheight{4pt}
\setlength{\tabcolsep}{2.0pt}
\setlength\arrayrulewidth{1pt}

\centering
\captionsetup{width=13.5cm}
\caption{Mean and median values of properties for comparison between GRQs and SRQs. Column (2): N refers to the number of sources. The p-values correspond to the K-S test and WMW test for the respective distributions. The comparisons of various properties of GRQs and SRQs are presented as follows:
I and II: lower and higher regimes of $z$ matched samples based on $z$ median as shown in Fig.\ \ref{fig:zmed1} and Fig.\ \ref{fig:zmed2}, respectively.}

\begin{tabular}{l c c c c c c c r}
\hline
Property & N &  Mean & Median & N & Mean &  Median &  K-S test  & WMW test  \\ 
 & & & & & & & p-value & p-value \\
 \hline
   \multicolumn{9}{c}{Redshift matched sub-samples based on z median}  \\ \hline
\hline
I. [ $z$ $\leq$ 1.00] & \multicolumn{3}{c}{GRQ} &   \multicolumn{3}{c}{SRQ}   &  \\ \hline
 
P$_{\rm core}$  [$\times$ 10$^{25}$ W Hz$^{-1}$] & 144 & 1.33 & 0.49  & 166 & 4.88 & 1.64 & 5.5 $\times$ 10$^{-16}$ & 2.9 $\times$ 10$^{-16}$ \\ 

CDF & 144 & 0.44 & 0.17 & 166 &  0.87 & 0.47  & 5.9 $\times$ 10$^{-9}$  & 1.3 $\times$ 10$^{-9}$ \\ 

P$_{\rm 1400}$  [$\times$ 10$^{26}$ W Hz$^{-1}$] & 173 & 1.92 & 0.78  & 166 & 3.10 & 1.64 & 9.4 $\times$ 10$^{-6}$ & 6.2 $\times$ 10$^{-7}$ \\

Q$_{\rm Jet}$  [$\times$ 10$^{44}$ erg s$^{-1}$] & 138 & 4.06 & 1.18  & 156 & 6.26 & 2.45  & 2.5 $\times$ 10$^{-5}$ & 1.5 $\times$ 10$^{-5}$ \\ 

$\rm \alpha^{\rm 1400}_{\rm 150}$ & 138 & 0.63 & 0.66 & 156 & 0.68 & 0.69 & 0.07 & 0.06 \\ 

M$_{\rm BH}$ [$\times$ 10$^{9}$ M$_{\odot}$] & 116 & 1.5  & 1.3  & 138 & 1.3 & 0.9 & 0.09 & 0.05 \\ 

$\lambdaup_{\rm Edd}$ & 116 & 0.06 & 0.04 & 138 & 0.09  & 0.06 & 4.2 $\times$ 10$^{-5}$ & 1.9 $\times$ 10$^{-4}$ \\ 

\hline

II. [1.00 < $z$ $\leq$ 2.45] & \multicolumn{3}{c}{GRQ} &   \multicolumn{3}{c}{SRQ}   &  \\ \hline
 
P$_{\rm core}$  [$\times$ 10$^{25}$ W Hz$^{-1}$] & 81 & 4.43 & 2.59 & 248 &  19.01  & 5.95 & 2.8 $\times$ 10$^{-7}$ & 4.2 $\times$ 10$^{-11}$ \\ 

CDF & 81 & 0.98 & 0.22 & 245 & 1.63  & 0.47  & 4.5 $\times$ 10$^{-4}$ & 6.2 $\times$ 10$^{-4}$ \\ 

P$_{\rm 1400}$  [$\times$ 10$^{26}$ W Hz$^{-1}$] & 89 & 5.05 & 3.56  & 245 & 12.26 & 6.16 & 2.8 $\times$ 10$^{-6}$ & 1.0 $\times$ 10$^{-7}$ \\

Q$_{\rm Jet}$ [$\times$ 10$^{45}$ erg s$^{-1}$] & 67 & 1.31 & 0.80 & 223 & 2.61  & 1.11  & 0.013 & 0.002 \\ 

$\rm \alpha^{\rm 1400}_{\rm 150}$ & 67 & 0.74 & 0.75 & 221 & 0.72 & 0.75 & 0.75  & 0.26 \\ 

M$_{\rm BH}$ [ $\times$ 10$^{9}$ M$_{\odot}$] & 75 & 2.7  & 1.9  & 227 & 2.3 & 1.6 &  0.17 & 0.06  \\ 

$\lambdaup_{\rm Edd}$ & 75 & 0.11  & 0.08 & 227 & 0.22 & 0.13 & 2.5 $\times$ 10$^{-4}$ & 2.5 $\times$ 10$^{-5}$ \\ \hline
\hline  
\label{zrpmed}
\end{tabular}
\end{table*}

%%%%%%%%%%%%%%%%%%%%%%%%%%%%%%%%%%%%%%%%%%%%%%%%%%%%

%%%%%%%%%%%%%%%%%%%%%%%%%%%%%%%%%%%%%%%%%%%%%%%%%%%%%

\begin{figure}
\includegraphics[scale=0.3]{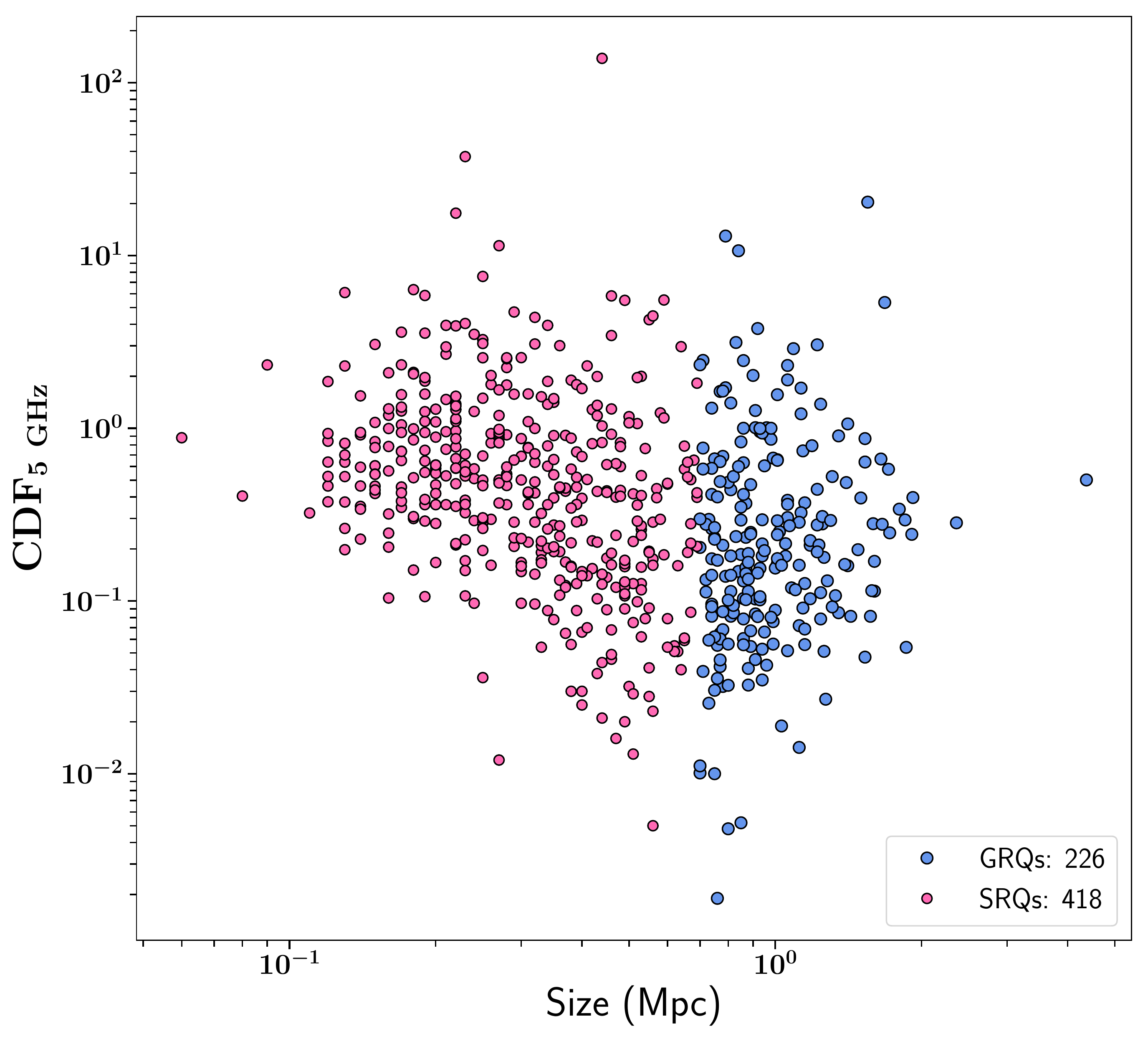}
\caption{This plot shows the core dominance factor (CDF) as a function of the size of radio-loud quasars. We observe a trend of decreasing CDF as the size of the source increases, possibly showing the effect of Doppler boosting.}
\label{fig:cdfsize}
\end{figure}
%%%%%%%%%%%%%%%%%%%%%%%%%%%%%%%%%%%%%%%%%%%%%%%%

\section{Discussion}
\label{Discussion}
In this section we discuss our results in detail and provide a possible model for growth of GRQs.
\subsection{CDF-size relation}
As per the relativistic beaming model \citep{blandford79}, the radio core emission is significantly enhanced with respect to the extended components when observed from an angle close to the line of sight in comparison to a larger viewing angle. Hence, CDF could be a statistical measure of the inclination of the jet axis with respect to the line of sight. As per the AGN orientation unification scheme, it is expected for the CDF to be anti-correlated with projected linear sizes of the sources i.e, sources with larger linear sizes should have less prominent cores as shown first in \citet{kapahisaikia82}. This is evident in the Fig.\ \ref{fig:cdfsize} where it can be observed that SRQs have more prominent cores as compared to GRQs. It is suggested in the literature (e.g. \citealt{Gopal1989}) that giant radio sources have higher core prominence as compared to smaller radio sources implying stronger nuclear activity. Our analysis shows that the CDF of SRQs is almost twice of GRQs in both bins of redshift matched samples. This is largely due to the beaming effect in which SRQs on average are viewed at smaller angles to the line of sight compared to GRQs.

%%%%%%%%%%%%%%%%%%%%%%%%%%%%%%%%%%%%%%%%%%%%%%%%%%%
 \begin{figure*}
 \centering 
\includegraphics[scale=0.34]{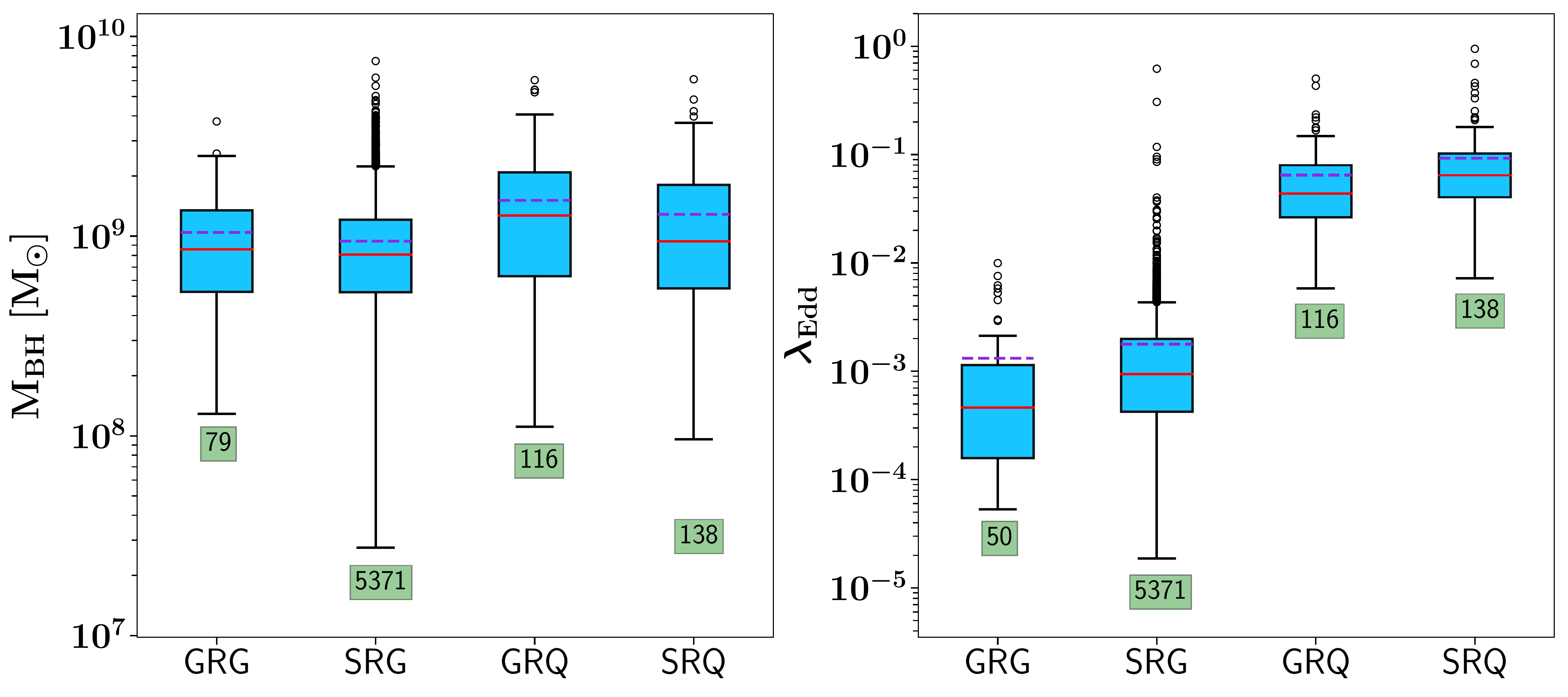}
\caption{The above two figures are box plots as described in Sec.\ \ref{sec:erall}, which shows distributions of black hole mass (left) and Eddington ratio (right) for GRGs, SRGs, SRQs, and GRQs. The number shown in the green box represents the number of sources in each sample. The red solid line indicates the median of the distribution and the dashed purple colour line denotes the mean.}
\label{fig:mbher}
\end{figure*}

%%%%%%%%%%%%%%%%%%%%%%%%%%%%%%%%%%%%%%%%%%%%%%%%%%%%%
 \begin{figure*}
 \centering 
\includegraphics[scale=0.28]{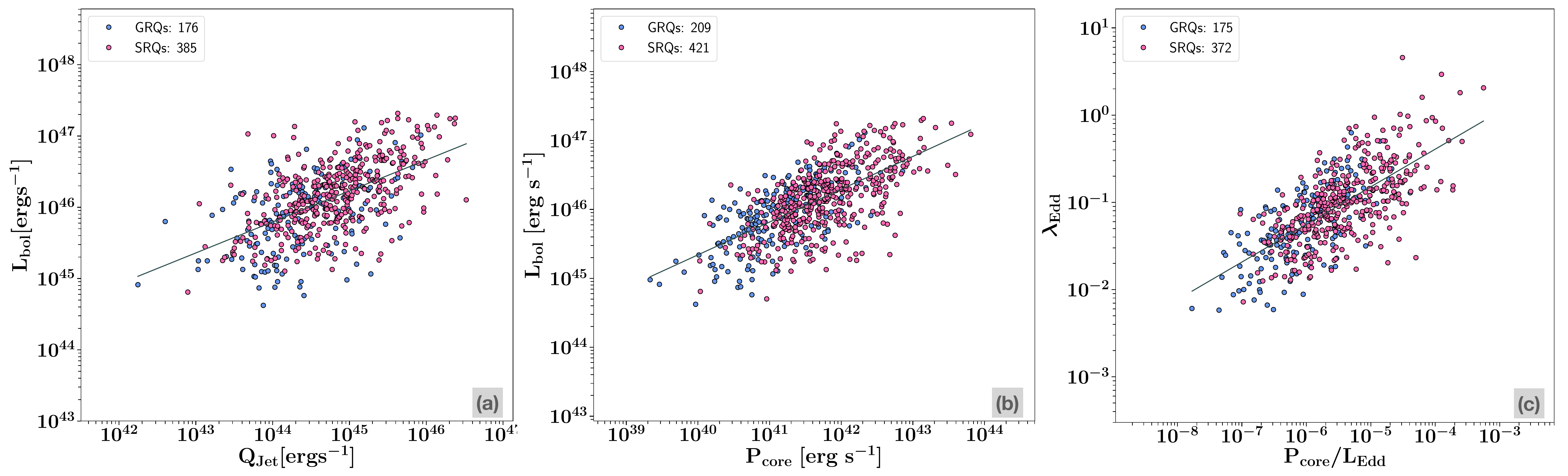}
\caption{Correlations between \lb and \qj in sub-figure (a), \lb and \pc in sub-figure (b), and \er and Eddington scaled \pc in sub-figure (c) for GRQs and SRQs. The Spearman's correlation coefficients are 0.55, 0.65 and 0.70 respectively.}
\label{fig:epqb}
\end{figure*}

%%%%%%%%%%%%%%%%%%%%%%%%%%%%%%%%%%%%%%%%%%%%%%%%%%%%%

\subsection{Role of black hole mass and Eddington ratio}\label{sec:erall}
In SAGAN.I paper, we presented the results on the \mbh and \er properties of giants with AGNs (non-quasars) having low as well as high excitation types, and it was found that the high excitation giant radio galaxies (HEGRGs) have higher \er than the low excitation giant radio galaxies (LEGRGs). A small overlap of \er was observed for HEGRGs and LEGRGs, which is also observed for the small-sized radio galaxies \citep{imogen18}, although they are thought to have bi-modal distribution as seen in the work of \citet{bh12rgs}. The \mbh distributions were found to be similar for SRGs and GRGs in SAGAN.I paper.

In order to investigate the role of \mbh and \er in the exceptionally large size of radio galaxies and quasars, we considered the samples from SAGAN.I paper in addition to our SRQ-GRQ samples for comparative study. To compare properties (\mbh and {$\lambda_{\rm Edd}$}) of GRGs, SRGs, GRQs, and SRQs we have opted to use box-plots, as seen in Fig.\ \ref{fig:mbher}, instead of the traditional histograms. 

The box-plots \citep{tukey1976exploratory} are also known as box and whisker diagrams or plots with outliers. Here, the upper and lower ends of the box represent the upper and lower quartiles respectively for a dataset. The two lines extending from the box are known as whiskers, which indicate variability outside the upper and lower quartiles. The dots are the outliers in the dataset, and the line in the middle of the box represents the median of the sample. These represent a graphical method that is better for showing variations in a data set and for analysis consisting of multiple datasets.

In Fig.\ \ref{fig:mbher}, we can see the box plot illustrating the distributions of \mbh and \er of GRGs and SRGs (non-quasar hosts) as well as GRQs and SRQs. In order to avoid possible problems with redshift evolution, we only considered objects matched in redshift, and hence, the sample is restricted below the redshift of 1. For \mbh we observe significant overlap between GRGs and GRQs, with GRQs tending to have higher \mbh than GRGs, and the giants having marginally higher values of \mbh than smaller sources for both galaxies and quasars. However, a clear bi-modal distribution or contrasting difference is observed for $\lambda_{\rm Edd}$, with negligible overlap. Also, this result indicates a higher accretion rate for the giants with quasar hosts or GRQs compared with GRGs; a similar result is seen for their smaller sized counterparts (SRGs and SRQs). Although the smaller sources appear to have marginally higher values of \er for both galaxies and quasars, it can be seen that \er does not depend strongly on the size of radio-loud AGNs. Now, having shown this result observationally, we can investigate other properties of SRGs and SRQs, and GRGs and GRQs which could provide us further clues for their giant sizes (e.g. environment). 

\subsection{Disc-jet coupling in radio quasars}
Extragalactic radio jets of various shapes and sizes have been known for nearly five decades \citep{hardcastlereview20}. To this date, new observations and numerical simulations reveal new information and help the quest for understanding the launching and collimation of astrophysical jets. The recipe of jet launching involves accretion, magnetic field, and spin but it is still not properly understood how an accretion disc is giving rise to ordered collimated outflows. According to the Blanford-Znajek mechanism \citep{bz77}, spin and magnetic flux threading the horizon determine the jet power. One of the possible ways to understand this is to explore the disc-jet symbiosis \citep{Falcke95,Willott1999,merloni03,fender10}. The disc-jet connection depicts an inter-link between the accretion process (rate, mode, field geometry and strength) and the power output or jets (\citealt{McNamara07}; \citealt{fender10}; \citealt{2016Stepanovs}), where the jets are rooted in the nuclear region of the accretion disc and move along the axis of rotation; a phenomenon also known as disc-jet coupling. One of the earliest observational evidences was given by \citet{1991Rawlings} through a correlation between jet power and disc luminosity indicator for FR-II radio galaxies. \citet{1999Cao} confirmed the same via probing the correlation between radio luminosity at 5 GHz and broad-line emission (an indicator for accretion power) for radio-loud quasars. Several other studies have also confirmed the inter-connection between disc-jet either via monitoring the X-ray and optical variability, and exploring radio - X-ray correlation for radio galaxy 3C120 (\citealt{marscherdiskjet02}; \citealt{ Ritaban09}) or through the correlation observed between broad-line region (BLR) luminosity and jet power traced using $\gamma$-ray luminosity for blazars and radio luminosity for RGs \citep{Sbarrato14}.

Recently, the Event Horizon Telescope (EHT) collaboration \citep{EHTmag} has imaged the polarised emission of the swirling plasma in the event horizon of the SMBH of Messier 87. The EHT observations of M87 along with favouring results from simulations indicate that the magnetic fields near the event horizon are dynamically important and play a crucial role in the formation of jets (\citealt{bz77}; \citealt{Narayan1994}; \citealt{Tchekhovskoy}; \citealt{2020Davis}) and in the physics (\citealt{Hawley1991};\citealt{2020Davis}) of the accretion disc. 

As the magnetic field lines encompassing the black hole build up near the event horizon, it gives rise to a magnetically arrested disc scenario (MAD; \citealt{Narayan2003}; \citealt{Tchekhovskoy}). It affects the disc properties and hinders the gas infall into the accreting black hole. The magnetic flux in the rotating disc is so intense that the magnetic field lines are piled up and twisted in the direction perpendicular to the plane of the accretion disc \citep{Zamani}, leading to the launching and collimation of jets. \citet{Zamani} have shown that magnetic field strength in jets strongly scales with accretion disc luminosity ({L$_{\rm bol}$}). Taking \lb as a proxy for magnetic flux in the jet-launching site, the {L$_{\rm bol}$}-\qj correlation indicates the contribution of the magnetic field in powering jets which in turn hints at the disc-jet coupling. Based on our results, we find strong evidence of disc-jet interplay for our sample of SRQs and GRQs (or RQs broadly speaking), as shown in Fig.\ \ref{fig:epqb} (a), where \lb representing accretion disc luminosity is plotted as a function of Q$_{\rm Jet}$. Spearman's correlation coefficient of 0.55 is supporting the scenario. Part of the scatter in the plot could be due to either measurement uncertainties or radiative loss due to the interaction of the jet with its immediate environment. Another factor could be the possible contribution of spin \citep{2013Velzen} apart from the accretion phenomenon contributing to jet kinetic power.

In SAGAN.I a similar correlation was found for a relatively smaller sample of GRGs confirming the close inter-link between disc properties and collimated jetted output of SMBH. We find that nearly 10\% of the accretion luminosity of GRQs and SRQs contributes to the jet kinetic power as given by the coupling parameter qj $\equiv$ Q$\rm _{\rm Jet}$/\lb = 0.09, which is quite less than the findings of \citet{Willott1999} who reported qj to be 20\%. Moreover, we state that our study sample has five SRQs and two GRQs with qj > 1. Based on the study by \citet{Ghisellini14}, qj > 1 is a plausible astrophysical scenario as they have shown for Flat Spectrum Radio Quasars (FSRQs) or BL Lac objects. They measured jet power from $\gamma$-ray luminosity and found that relativistic jets have more power than the luminosity of the accretion disc. Such a scenario can be explained when the extraction of spin energy from a black hole \citep{Tchekhovskoy} is invoked along with disc-jet coupling. The higher spin values for jet production in radio-loud quasars has also been suggested by other studies \citep{Schulze2017,chen21}.

In addition, we have also examined the possible correlation between \lb and \pc(Fig.\ \ref{fig:epqb} b), and found them to have a strong correlation with a Spearman's correlation coefficient of 0.65. When normalised to L$_{\rm Edd}$, the resulting correlation of \er as a function of P$_{\rm core}$/L$_{\rm Edd}$ even gets stronger with a correlation coefficient of 0.7 (Fig.\ \ref{fig:epqb} c). Due to the resolution limit (5\arcsec), the radio emission of the nuclear region (or radio core) that we are addressing here possibly includes parsec-scale jets in some sources. Therefore, this relation provides another piece of evidence for the contribution of the accretion properties of the SMBH in fueling the relativistic radio jets. Fig.\ \ref{fig:epqb} with its sub-plots gives a clearer view of how the disc magnetic field affects the formation of relativistic jets. The high Eddington ratio (accretion rate) leads to a higher accumulation of magnetic fields in the vicinity of the black hole and once the magnetic field reaches the saturation state, the magneto-rotational instabilities are triggered. This process carries away the angular momentum of the disc into a jetted outflow. The spin plays the role of twisting the magnetic field lines and confirms the transfer of magnetic energy into the Poynting flux of relativistic jets.

Lastly, there have been studies suggesting the possible contribution of relativistic beaming affecting the optical continuum emission in quasars. However, recent studies of \citet{vangorkom15} have shown that such contribution is negligible.

\subsection{SRQ to GRQ}
\label{GRQ-RQ}
In this decade, the arrival of new sensitive instruments and advancement in software has facilitated the discovery of a larger sample of GRGs, especially the ones with low surface brightness (e.g. \citealt{PDLOTSS}; \citealt{Delhaize21}). However, the question about the conditions under which an SRG or SRQ becomes a giant remains to be well understood. In an effort to answer this question we have carried out a comparative study between the properties of GRQs and SRQs. In Sec.\ \ref{sec:prop}, distributions of properties concerning the host and large scale properties of GRQs and SRQs have been discussed. Our results show that SRQs have more prominent cores as compared to GRQs. They have \pc and CDF twice as that of GRQs. Similarly, for large scale properties like \P and Q$_{\rm Jet}$, SRQs tend to show more powerful jets carrying almost double the radio power than the GRQs. However, interestingly, irrespective of SRQs appearing to be more powerful, the spectral index distributions of both SRQs and GRQs are observed to be the same. It indicates that the AGN activity in the GRQs has not ceased. Owing to their larger sizes, GRQs are more susceptible to higher radiation losses due to adiabatic expansion and inverse Compton scattering and yet have similar spectral indices between about 150 and 1400 MHz. This indicates that perhaps the AGN activity is more dominant or prolonged than SRQs. Note that re-acceleration of radiating particles in the jets and lobes can also lead to flatter spectral indices.

If we consider the samples in the lower and upper redshift bins, it is observed that sources at high redshifts are more powerful. This is because with the increase in redshift there is a decrease in detection of less powerful sources due to Malmquist bias. It is also noticed that sources in the upper bin (high $z$) have steeper spectra. It could possibly be the consequence of radio luminosity - spectral index (P-$\alpha$) correlation where sources with high radio power are observed to have steeper spectral indices (\citealt{Blundell}; \citealt{DabhadeSAGAN20}). Sources with high \qj have enhanced magnetic fields in the hotspots. It eventually results in the rapid cooling of relativistic electrons,  and as a result, electrons with steeper spectra are injected into the lobes.

The black hole mass distributions of both SRQs and GRQs are found to be similar statistically, although, the former are accreting faster than the latter, which is reflected from their higher values of \er as compared to GRQs. Based on our analysis, we suggest the following basic model for the growth and evolution of GRQs: the GRQ progenitor is an SRQ with a high accretion rate, which grows with time in terms of overall size. This is supported by the self-regulated black hole growth scenario \citep{Hopkins2005a,Hopkins2005b}, where it is shown that the feedback energy from the AGN can heat up gas in the host galaxy and result in quenching of the accretion process. This in turn will lead to a decrease in the $\lambda_{\rm Edd}$. Hence, this could lead to reduction in the overall output from the disc and affect the jet power. Our results presented in Sec.\ \ref{sec:prop} show that P$_{\rm core}$, Q$_{\rm Jet}$, P$_{1400}$, and \er are less for GRQs compared with SRQs, which is consistent with the above model. This can be tested by comparing the `effective' lifetimes \citep{Hopkins09} of the quasars with that of the spectral and dynamical ages of the SRQs and GRQs. The measurement of spectral and dynamical ages of large samples of SRQs and GRQs via multi-frequency radio observations is quite difficult owing to the large time requirements. Spectral age measurements of a few quasars \citep{parma99} indicate their ages to be in the range of $\sim$\,1 to 20 Myr which is comparable to the quasar lifetimes obtained in the studies of \citet{Hopkins2005a}.

The signs of restarted radio AGNs \citep{saikiaddrgs} based on radio morphology were revealed with the discovery of `double-double radio galaxies' \citep{Schoenmakersddrgs} nearly two decades ago with only about 110 such sources known so far. However, only a very few have quasar hosts (e.g. 4C02.27 or J0935+0204-\citealt{jamrozy09}, J0746+4526-\citealt{nandi14}, J1244+5941-\citealt{saikiaddrgs}) indicating their rarity. Also, there are only a few examples of RQs with intermittent jet activity (e.g. PKS 1127-145-\citealt{Siemiginowska07}; 3C273-\citealt{Stawarz04}). Hence, it is possible that SRQs and GRQs have episodic activity on a very short time scale, which seldom manifests into the double-double morphology for us to observe.

\subsection{High redshift GRQs}
The radio emission from radio sources is quenched due to enhanced inverse Compton losses against the cosmic microwave background for higher redshift sources. For a radio-loud AGN which has grown to larger volumes over a period of a relatively long time, the internal magnetic energy density in the radio lobes is expected to be less than the energy density of the CMB photons. This was clearly demonstrated by \citet{Ishwara1999} for their sample of giant and smaller sources. In such scenarios, the relativistic radio-emitting electrons will preferentially lose energy via inverse Compton collisions with the CMB photons. This will lead to enhanced X-ray emission and suppression of radio emission. The hotspots in the lobes are relatively less affected by this due to their compactness and higher magnetic energy density. There is a (1+$z$)$^{4}$ factor dependence of the CMB energy density, and hence, for high redshifts, the quenching of radio emission is expected to be much larger. Therefore, at higher redshift, inverse Compton ghosts of radio galaxies or radio quasars \citep{icghost11} are observed, where the jets of such sources have switched off and the lobes are visible in X-ray only (e.g. \citealt{Tamhane2015}).

In our GRQ-catalogue we have eight high-$z$ sources with 2 $<$ \z $\leq$ 2.94, having radio powers \simi 10$^{27}$ \wz at 1400 MHz. The radio maps reveal their FR-II radio morphologies, indicating that these GRQs must be enormously powerful to negate the quenching effect of high-$z$ CMB. Studies \citep{icghost11,fabian14,Ghisellini15} have shown the existence of double-lobed radio sources at higher redshifts, however not for sources of megaparsec sizes. Recently, \citet{Banados21} discovered the farthest radio-loud quasar at the redshift of 6.82, which is when the age of the Universe was only \simi0.8 Giga years. These studies have shown the existence of radio jetted AGNs at higher redshifts or when the Universe was at its early stages, where black holes managed to grow to supermassive sizes. It is possible that a population of such sources is currently invisible to us due to the sensitivity limits, and hence, are prime targets for future telescopes like the Square Kilometre Array (SKA) and \textit{Athena}. Based on the best current data (radio and optical) available, only three of the total 200 quasars known above the redshift of 6 are radio-loud \citep{Banados21}. High-$z$ radio-loud AGNs are often found in over-dense environments, and hence, also are tracer of protoclusters \citep{Venemans07,Wylezalek13, hatch14} which could be the progenitors of the present-day galaxy clusters. It is vital to study these objects as they are probes of high redshift Universe, and their studies hold the key to understanding the formation, growth and evolution of radio sources.

\section{Summary}
\label{Summary}
All the results of our study and analysis are summarised as follows:
\begin{itemize}
    \item We report the discovery of 10 new GRQs in the redshift range of 0.66 to 1.72 with the largest one having the projected linear size of 1.55 Mpc. All of them have FR-II type radio morphology. 
    
    \item We have constructed the GRQ catalogue consisting of 121 sources from \citet{DabhadeSAGAN20}, 134 sources from \citet{Kuzmicz2021}, and 10 new GRQs. We also created a well-defined SRQ sample of 422 sources selected from \citet{Kimball}. 
    
    \item To understand how similar GRQs and SRQs are, we have compared their various properties like P$_{\rm core}$, CDF, P$_{1400}$, Q$_{\rm Jet}$, $\alpha_{150}^{1400}$, \mbh and $\lambda_{\rm Edd}$. For samples in the redshift-matched bins, we found that SRQs have higher P$_{\rm core}$, CDF, P$_{1400}$, Q$_{\rm Jet}$, and \er as compared to GRQs. But in the case of M$_{\rm BH}$ and $\alpha_{150}^{1400}$, they both show similar distributions. The higher values of CDF suggest that statistically, the SRQs are inclined at smaller angles to the line of sight than the GRQs.
    
    \item Our results show that the central engines of SRQs are more powerful than those in GRQs.
    
    \item When compared to GRGs and SRGs, the GRQs and SRQs have higher \mbh and \er indicating quasar engines to be more active. 
    
   \item We find tight correlations between \lb and Q$_{\rm Jet}$, and \lb and P$_{\rm core}$. This implies that the disc-jet coupling has a significant contribution towards launching jets. 
   
   \item The \er distributions of both GRGs and GRQs, and SRGs and SRQs are bimodal with the quasars having significantly higher values of $\lambda_{\rm Edd}$. The smaller sources appear to have marginally higher values of \er for both galaxies and quasar suggesting only small changes in the accretion process as radio sources grow to their giant sizes.
   
\end{itemize}

\section*{Acknowledgements}
We thank the anonymous reviewer for critically reading the manuscript and providing with valuable comments. We thank IUCAA (especially Radio Physics Lab\footnote{\url{http://www.iucaa.in/~rpl/}}), Pune for providing all the facilities during the period the work was carried out. LCH was supported by the National Key R$\&$D Program of China (2016YFA0400702) and the  National Science Foundation of China (11721303, 11991052). We gratefully acknowledge the use of Edward (Ned) Wright's online Cosmology Calculator. This research has made use of the VizieR catalogue tool, CDS, Strasbourg, France \citep{vizier}.
This research has made use of the CIRADA cutout service at URL\footnote{\url{http://cutouts.cirada.ca}}, operated by the Canadian Initiative for Radio Astronomy Data Analysis (CIRADA). CIRADA is funded by a grant from the Canada Foundation for Innovation 2017 Innovation Fund (Project 35999), as well as by the Provinces of Ontario, British Columbia, Alberta, Manitoba and Quebec, in collaboration with the National Research Council of Canada, the US National Radio Astronomy Observatory and Australia's Commonwealth Scientific and Industrial Research Organisation.

We acknowledge that this work has made use of  \textsc{astropy} \citep{astropy}, \textsc{aplpy} \citep{apl}, \textsc{matplotlib} \citep{plt} and \textsc{topcat} \citep{top05}.

%%%%%%%%%%%%%%%%%%%% REFERENCES %%%%%%%%%%%%%%%%%%
\bibliographystyle{aa} 
\bibliography{SAGAN3_GRQS.bib}
%%%%%%%%%%%%%%%%%%%%%%%%%%%%%%%%%%%%%%%%%%%%%%%%%%

% %%%%%%%%%%%%%%%%% APPENDICES %%%%%%%%%%%%%%%%%%%%%
%\appendix  

%%%%%%%%%%%%%%%%%%%%%%%%%%%%%%%%%%%%%%%%%%%%%%%%%
\end{document}